# Modelling of the Fender Bassman 5F6-A Tone Stack

Dr Steven Fenton
School Of Computing & Engineering
The University Of Huddersfield
Queensgate, Huddersfield, UK
s.m.fenton@hud.ac.uk

*Abstract*—This paper outlines the procedure for the effective modelling of a complex analogue filter circuit. The Fender Bassman 5F6-A is a circuit commonly employed in guitar amplifiers to shape the tonal characteristics of the amplifier output. On first inspection this circuit may look rather simple, however the controls are not orthogonal, resulting in complicated filter coefficients as the controls are varied. This in turn can make the circuit difficult to analyse without the use of mathematical emulation tools such as PSPICE or MATLAB. First the circuit is described, a method of analysis is proposed and general expressions for continuous-time coefficients are given. A MATLAB model is then produced and the frequency responses of which are shown.

I. Introduction

This motivation behind this paper is primarily to provide a complete analysis of the 5F6-A tonestack, with a focus on the analysis techniques utilised and resulting MATLAB code. It should be of benefit to readers who are looking to emulate this circuit or similar circuits as the techniques employed are well established and can be employed elsewhere.

II. Description of the Circuit

A. *Overview*

Commonly found in many guitar amplifiers, especially those that derive from the Fender design, the tone stack filters the signal of he guitar in a unique and non-ideal way. The user can adjust Treble, Middle, and Bass controls to modify the gain of the respective frequency bands. However, these controls are not orthogonal, and changing some controls affects the other bands in a complex way.

The full Bassman schematic can be found online [1] and in guitar amplifier books. While other guitar amplifiers may vary, the tone stack in the Bassman topology is found after the preamplifier stages and before the phase splitter. It's highly likely that you will also find that the tone stack is preceded by a cathode follower in order to isolate the tonestack itself and reduce loading effects. Typically this presents a 1kΩ load to the input and the phase splitter stage presents a 1MΩ load to the output.

The Fender '59 Bassman tone stack circuit is shown in Fig. 1. The Treble, Middle, and Bass knobs are potentiometers, which have been modeled here as parameterized resistors, $R_t$, $R_m$ and $R_b$ respectively. The Treble and Middle controls use linear potentiometers, while the Bass control uses a logarithmic taper potentiometer. In this paper, t and m correspond to the Treble and Middle controls and range in value from [0, 1]. The Bass control, b, also ranges from [0, 1].

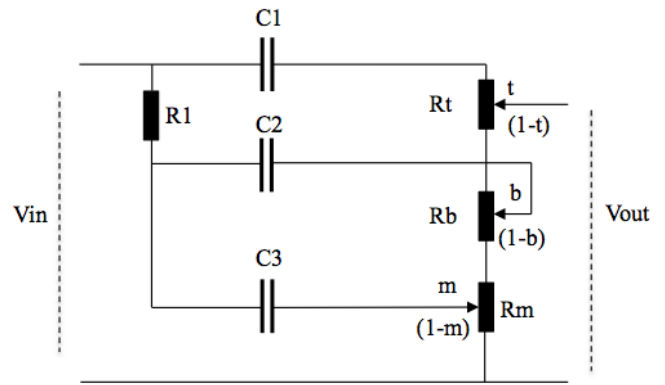

Fig. 1. Fender tonestack circuit.

B. *Non-orthogonal Controls and Approximation*

The Bassman circuit has three potentiometers which offer control over the Treble, Mid and Bass range of frequencies. With reference to Fig 1, it can be seen that the corner frequency of the filter formed by C1 and Rt is affected by the position of the Rb potentiometer. Similar interaction is caused by the position of Rt which determines the Vout mix between the voltage present across C1 and that present across Rb & Rm.

Visualising the circuit and re-drawing as three separate circuits can approximate the circuit output [3]. In this case, the interaction between the three bands caused by the potentiometers is ignored. Whilst this can give a good overview of circuit operation, it is not ideal should full modeling of the circuit operation be required.

## III. METHOD OF ANALYSIS

In order to analyse the circuit operation fully, taking into account its non-orthogonal operation, Kirchoff's circuit laws can be applied. Using Kirchhoff's circuit laws, one can either do nodal analysis using Kirchhoff's current law (KCL) or mesh analysis using Kirchhoff's voltage law (KVL) or a combination of the two.

Nodal analysis writes an equation at each electrical node (points in the circuit where elements or branches connect), requiring that the branch currents incident at a node must sum to zero. The branch currents are written in terms of the circuit node voltages. By adopting this method, voltages between nodes can be determined anywhere on the circuit where the branch current has been derived and the resistance or impedances are known.

Using mesh analysis, the voltages around each closed loop can be expressed in terms of relative impedances and currents around that loop. Referring to Fig 2, the loops are I1, I2 and I3. For basic use of mesh analysis, the following list gives a short step-by-step instruction to create the mesh matrix:

1. Convert all current sources to voltage sources and redraw the circuit using these voltage sources
2. Select a mesh current variable for each loop, e.g. I1, I2, ..., I$N$ where $N$ is the number of loops.
3. From KVL it follows that the matrix representation will have the general form of

$$\begin{bmatrix} Z_{11} & .. & Z_{1N} \\ \vdots & \vdots & \vdots \\ Z_{N1} & .. & Z_{NN} \end{bmatrix} \begin{bmatrix} I_1 \\ \vdots \\ I_N \end{bmatrix} = \begin{bmatrix} V_1 \\ \vdots \\ V_N \end{bmatrix}$$

(1)

$Z_{NN}$ in each case is the overall complex impedance forming the mesh.

Equation (1) can be expanded and by inspection of Fig 2 in terms of relative impedances and loop currents, the following three equations can be derived.

$$I_1[R1 + Rm_{1-m} + XC_3] - I_2 R1 - I_3 XC_3 = Vin$$

$$-I_1 R1 + I_2[Rt + R1 + XC_2 + XC_1] - I_3 XC_2 = 0$$

$$-I_1 XC_3 - I_2 XC_2 + I_3[Rb_{1-b} + Rm_m + XC_2 + XC_3] = 0$$

(2)

Where $XC_x$ represents the reactance of the associated capacitor in the circuit.

In addition, given Kirchoff's law that all voltage sources forming a mesh must sum to zero, known values may be substituted into the equations. It is then possible to solve (usually by simultaneous equation methodology) the unknown voltages at each node.

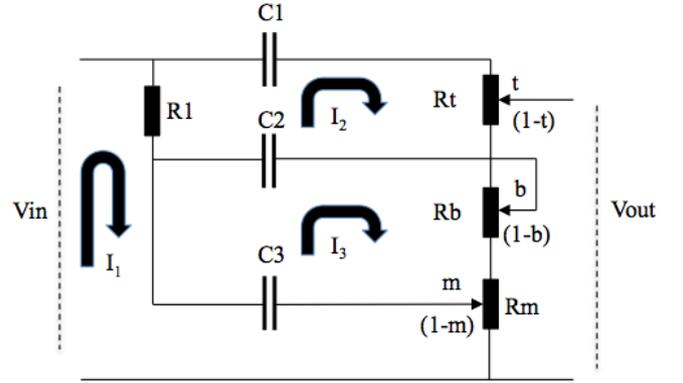

Fig. 2. Tonestack current loops.

Alternatively, the equations can be solved through the use of MATLAB as follows.

## IV. MATLAB IMPLEMENTATION

Equation (1) can be expressed in terms of admittance as

$$[V_n][Y_{nN}] = [I_n]$$

(3)

Where $Y_{nN} = [Z_{nN}]^{-1}$ (4)

In MATLAB, we can create two matrices, one forming an impedance matrix $Y$ and another forming the unknown loop currents. An additional matrix can be formed based on the known voltages around each closed loop. In order to plot a response of *Vout* with respect to *Vin*, an arbitrary voltage can be chosen as *Vin* as a starting point.

In the code fragment shown in Fig 3, *Xcn* represents the reactance of the associated capacitor $n$. This will be frequency dependent and therefore a 'for-loop' can be incorporated to enable calculation of each loop current versus frequency. The code shown indicates how to represent the impedances, in this case relating to *Vin* in (2).

```
Xc3 = (1/(2*pi*F*c3));
…
z1 = complex(r1+rm2, Xc3);
z2 = -r1;
z3 = complex(0, -Xc3);
…
Y = [z1 z2 z3;
     z4 z5 z6;
     z7 z8 z9;];
```

Fig. 3.  Tonestack MATLAB code fragment.

We can compute [I] by using the command

```
I = V * inv(Y)
```

Where `inv(Y)` is the inverse of matrix Y, thus representing an admittance matrix detailed in (4). Once the mesh currents have been solved at each frequency, voltages at every point of the circuit can be resolved through ohms law.

## V.  MODELLED RESPONSES

The following plots show the modelled response of the tonestack using the full code outlined in Fig 8. Note the characteristic mid range scoop evident in all plots, even with the mid range parameter set to minimum. Fig 7 shows incremental settings of each parameter in isolation for comparative purposes. The parameters (b, m & t) have been swept in 0.1 steps from 0 to 1 in each case.

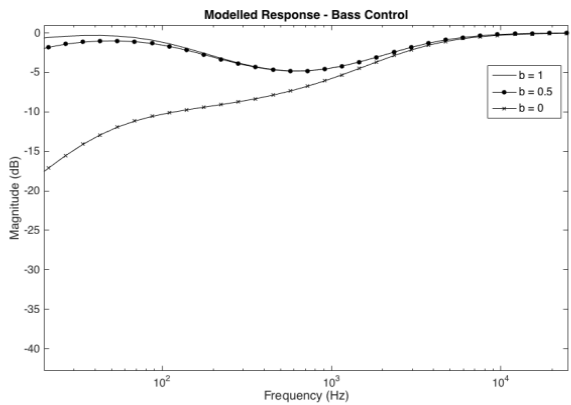

Fig. 4.  Modelled response with bass parameter changing.

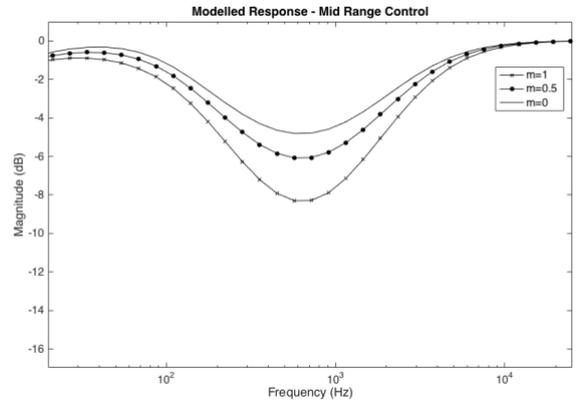

Fig. 5.  Modelled response with mid range parameter changing.

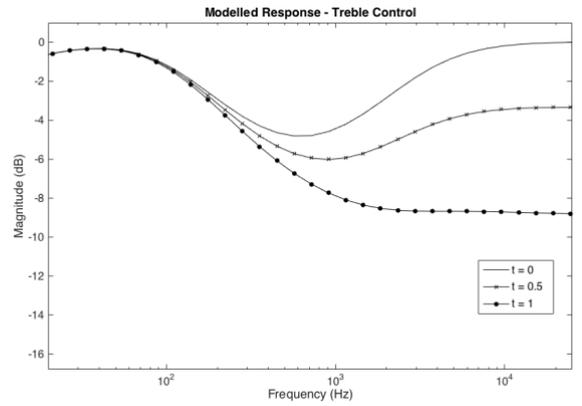

Fig. 6.  Modelled response with treble parameter changing.

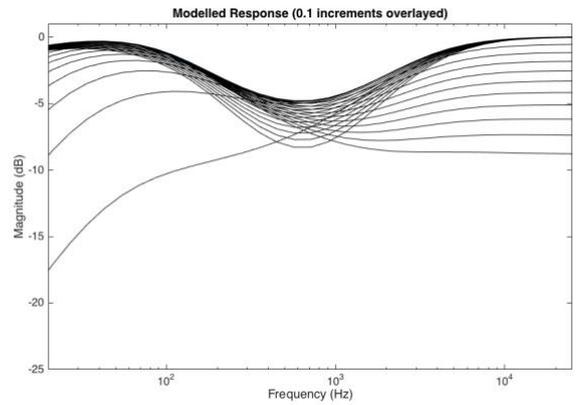

Fig. 7.  Modelled responses showing 0.1 increments on each parameter in isolation.

# VI. MATLAB SCRIPT

```
%--------------------
% Filename
% Tonestack_mesh_SOLUTION.m
% Date - 21st December 2015
% Author - Steven Fenton
% Contact - s.m.fenton AT hud.ac.uk
%
% This Matlab Script is a model of the
% Fender Bassman 5F6-A Tonestack
% There are some approximations of this
% Tonestack available on the Web
% and also a really great analysis can be
% found in the excellent book by
% Richard Kuehnel.
% Despite this, the analyses don't
% define complex impedance solutions.
% It's important to do this in order to
% get both a true magnitude response of
% the tonestack in addition to input and
% output phase responses.
% This script represents a complex
% impedance solution.
%
% You can modify component values as.
% r1 = 56k (ohms)
% c1 = 220*10^-12 (farad)
% c2 = 0.022*10^-6 (farad)
% c3 = 0.022*10^-6 (farad)
% rt = 220k (ohms)
% rm = 25k (ohms)
% rb = 1M (ohms)
%
% b, m and t are coefficients
% representing the wiper
% positions of the three potentiometers.
%--------------------

clear all;

% Component  Values of Tone Stack
r1 = 56000;
rt = 220000;
rm = 25000;
c1 = 220*10^-12;
c2 = 0.022*10^-6;
c3 = 0.022*10^-6;
rb = 1000000;
%--------------------

% Wiper positions
t = 0; % 0->1 0.5 being centre tap
       % 1 constitutes maximum treble cut
m = 0; % 0->1 0.5 being centre tap
       % 1 constitutes maximum mid cut
       % N.B. there will always
       % be a characteristic mid scoop.
b = 1; % 0->1 0.5 being centre tap
       % 0 constitutes maximum bass cut
%--------------------

% Calculate resistances based on wiper positions
rt1 = (rt*t); %output voltage pot leg
rt2 = (rt*(1-t));
rm1 = (rm*m); %Middle next to bass pot
rm2 = (rm*(1-m)); %Bottom, next to ground
rb1 = 1000000 * b;
%--------------------

% Arbitrary voltage level of input
Vin = 5;

% Create logspaced frequency
xspace = logspace(0,5, 50);
plotsizes = length(xspace);

FSpace = zeros(plotsizes, 1);
for freq = 1:length(xspace)
  FSpace(freq) = xspace(freq);
end

% Store actual frequency values
% for ease of plotting
xFrequencies = FSpace';

% Mesh current storage
plot_Iout1 = zeros(plotsizes, 1);
plot_Iout2 = zeros(plotsizes, 1);
plot_Iout3 = zeros(plotsizes, 1);
% Vout Storage
Vout = zeros(plotsizes, 1);

% Loop through all frequencies to
% calculate mesh currents
for freq = 1:length(xspace)

  F = xspace(freq);
  % Reactance Calculations
  Xc1 = (1/(2*pi*F*c1));
  Xc2 = (1/(2*pi*F*c2));
  Xc3 = (1/(2*pi*F*c3));

  % I(1)
  z1 = complex(r1+rm2, Xc3);
  z2 = -r1;
  z3 = complex(0, -Xc3);
  % I(2)
  z4 = -r1;
  z5 = complex(rt1+rt2+r1, (Xc2+Xc1));
  z6 = complex(0, -Xc2);
  % I(3)
  z7 = complex(0, -Xc3);
```

```
    z8 = complex(0, -Xc2);
    z9 = complex(rb1+rm1, (Xc2+Xc3));

    Y = [z1 z2 z3;
         z4 z5 z6;
         z7 z8 z9;];

    % voltage vector is entered as a
    % transpose of row vector
    V = [5, 0,0]';

    % Calculate loop currents
    I = inv(Y)*V;

    % Store loop currents per frequency
    plot_Iout1(freq) = I(1);
    plot_Iout2(freq) = I(2);
    plot_Iout3(freq) = I(3);

end %for freq loop

% Voltages across components
Vrm2 = rm2.*abs(plot_Iout1);
Vrt1 = rt1.*abs(plot_Iout2);
Vrt2 = rt2.*abs(plot_Iout2);
Vrb_rm1 = (rb1+rm1).*(abs(plot_Iout3));

% Calculate voltage output as the sum of
% voltages formed by the product of
% relative impedances and associated
% mesh currents

VTotal = (Vrm2)+(Vrt2)+(Vrb_rm1);
Vout_dB = 20*log10(VTotal/Vin);
%---------------------

% Create Plot
figure(1);
semilogx(xFrequencies, Vout_dB, '-s');
axis([20, 24000, -inf, 1]);
ylabel('dB')
xlabel('Hz');
title('Modelled Response');
%---------------------
```

Fig. 8. Complete Matlab script for Fender tonestack modelling

REFERENCES

[1] Ampwares, "5F6-A schematic," Retrieved June 23rd, 2016, [Online] http://ampwares.com/schematics/59_bassman_manual.pdf
[2] R. Kuehnel, Circuit Analysis of a Legendary Tube Amplifier: The Fender Bassman 5F6-A, 2nd ed. Seattle: Pentode Press, 2005.
[3] Ampbooks, "Fender Bassman Tonestack Low Frequency Response," Retrieved July 5th, 2016, [Online] https://www.ampbooks.com/mobile/classic-circuits/bassman-tonestack-low-frequency/